\documentstyle[sprocl,epsf]{article}

\def\ra{\rightarrow}

\def\beq{\begin{equation}}
\def\eeq{\end{equation}}
\def\bea{\begin{eqnarray}}
\def\eea{\end{eqnarray}}

\def\beq{\begin{equation}}
\def\eeq{\end{equation}}
\def\bdm{\begin{displaymath}}
\def\edm{\end{displaymath}}
\def\bea{\begin{eqnarray}}
\def\eea{\end{eqnarray}}

\def\vg{\gamma^{*}}                                     

\def\D0{D\0~}

\def\ra{\rightarrow}

\begin{document}

\begin{flushright}
MSUHEP-00615\\
hep-ph/0006176\\
\vspace{2.5cm}
\end{flushright}
\title{SOFT PARTON RESUMMATION IN THE CURRENT REGION OF  
SEMI-INCLUSIVE DEEP INELASTIC SCATTERING\footnote{Presented at the 8th
Intl. Workshop on Deep Inelastic Scattering (DIS2000), Liverpool,
U.K., April 2000}}

\author{\underline{P. NADOLSKY}{}$^{(a)}$, 
D. R.  STUMP{}$^{(a)}$, C.-P. YUAN{}$^{(b)}$\footnote{
On leave of absence from Michigan State University.}}

\address{
$(a)$ Department of Physics \& Astronomy, Michigan State University, 
E. Lansing,\\ MI 48824, U.S.A. \\
$(b)$ Theoretical Physics Division, CERN, CH-1211 Geneva 23, Switzerland
}

\maketitle\abstracts{We discuss resummation of large logarithmic terms
that appear in the cross-section of semi-inclusive
DIS in the case when the final-state hadron 
follows the direction of the incoming electroweak vector boson
in the c.m.~frame of the vector boson and the initial-state proton. 
}

During the past years, the H1 and ZEUS
Collaborations at DESY-HERA have put a significant effort\,\cite{desy99091} 
into the experimental
study of  semi-inclusive deep 
inelastic scattering (sDIS) $e p \ra e B X$. 
In this process, in addition
to the scattered electron (or positron) 
$e$, some specific final-state hadron $B$, 
or the flow of the hadronic energy into a given region of phase
space, is observed. The produced data set 
includes events  with a large momentum
transfer from the electron to the hadronic system, $Q^2 = - q^2 \gg 
\Lambda_{QCD}^2$. 
The theoretical
description of such events can be attained with the help of 
methods of perturbative QCD, 
under the assumption of a single-photon exchange
between the scattered electron and the proton. If the  
renormalization and factorization scales in the 
calculation are chosen to be of the order of 
the large photon virtuality $Q$,  then the cross-section can be calculated
perturbatively as a series in the small QCD coupling 
$\alpha_s$. However, the smallness
of $\alpha_s$ does not guarantee fast convergence of this
series. Even when there is a large momentum scale in the collision,
it is not hard to imagine a situation in which the 
sDIS observable  receives important contributions from all orders of
the perturbative expansion.

\begin{figure}[t]
\epsfysize 4.8cm
\epsffile{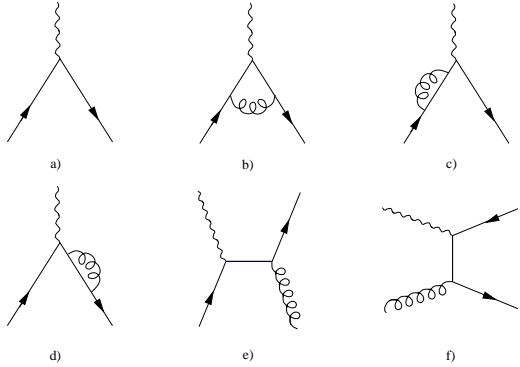}
\caption{The diagrams contributing to sDIS at the next-to-leading order}
\end{figure}

To illustrate this point, consider 
semi-inclusive DIS in the c.m.~frame of the photon and the 
initial proton. The $z$-axis in this frame is defined to follow the
direction of the photon's momentum.  We assume all particles to 
be massless.

At the order ${\cal O}( \alpha_s^0)$ (Fig.\,1a), the photon
interacts with an initial-state quark $a$ whose momentum is 
\beq
\vec p_a = x \vec P_A = - x \vec q \ ; \qquad 0 \leq x \leq 1.
\eeq
The final-state quark $b$, which has momentum  equal to
$\vec p_a + \vec q= ( 1 - x) \vec q $, escapes in the direction of 
motion of the incoming photon, {\it i.e.}~along the positive 
$z$-axis. At the order ${\cal O}( \alpha_s)$, 
one needs to account for QCD  loop cor\-rec\-tions (Figs.\,1b-d), 
as well as for diagrams with the emission of an additional parton 
(Figs.\,1e,f).
The individual diagrams may contain soft singularities,
but these cancel in the coherent sum of all contributions.
The remaining collinear singularities are absorbed
into the parton distribution functions (PDFs), 
or fragmentation functions (FFs).

While the above computational procedure yields a finite result, it
fails to describe accurately the rate of  sDIS in the limit when
the direction of the resolved final-state parton is close to the one 
predicted in the leading order subprocess, that is in the direction of
the initial electromagnetic current. Mathematically, the higher order
corrections in this kinematic region (the current region)
 of sDIS are dominated by large 
logarithmic terms $\log^m q_T / Q$. 
Here $q_T$ is a variable that is related to the final hadron's
pseudorapidity $ \eta$ in the $\vg p$ c.m.~frame:
\beq
q_T = W e^{-\eta}, 
\mbox{ with } W^2 = Q^2 \left(\frac{1}{x}-1 \right). \label{qt}
\eeq
The leading order process has $\eta =
+\infty$, {\it i.e.}~$q_T = 0$.  The singular 
logarithmic terms in the limit $q_T \ra 0$
are therefore given by 
\beq
\frac{1}{q_T^2} \sum_{n=1}^{\infty} \alpha_s^n \sum_{m=1}^{2 n - 1 } 
v^n_m 
\log^m \frac{q_T}{Q}~,
\label{singsum}
\eeq
where $v^n_m$ denote coefficients of order 1.
As can be seen from (\ref{singsum}), 
due to the presence of large logarithms, the higher order
contributions are not negligible when $q_T$ is small.
Thus, they all must be accounted for in order
to predict reliably the observables in this kinematical regime.

It is interesting to note that a structure of the singular terms that
is similar to the all-order sum (\ref{singsum}) 
also dominates the cross-section
for the production of back-to-back hadronic
jets at $e^+ e^-$ colliders \cite{CS}, and the cross-section for  
small-$p_T$ vector boson production at hadron colliders \cite{CSS}. 
For those two processes, it has been proven that the analog of
(\ref{singsum}) can be computed in a closed form,  
by considering a Fourier transform of the cross-section from 
momentum space to the space of the conjugate variable 
(the impact parameter) $\vec b$ \cite{PP,CS,CSS}. 
The result is called the resummed cross-section. 
Crossing relations between sDIS, $e^+ e^-$
hadroproduction and vector boson production suggest that the same
resummation technique may be successful in improving the theoretical 
description of the sDIS observables.
Indeed, it was shown \cite{MOS} that the $b$-space resummation 
formalism provides a description of
the energy flow in the small-$q_T$ region of sDIS. 
This approach was generalized \cite{NSY} 
to describe particle production cross-sections and
multiplicities, and a comparison with experimental data was presented.

In the $b$-space formalism \cite{CSS}, 
the resummed cross-section in the limit $q_T \leq Q$ 
is written in the form
\beq
\sigma (e A \ra e B X)  \approx N^\prime 
\int \frac{d^2 b}{(2\pi)^{2}} e^{i\vec q_T \cdot
\vec b } \tilde W_{BA}(b,Q) + \sigma^{pert}_{BA} - \sigma^{sing}_{BA}, 
\label{resum}
\eeq
where
\beq
\tilde W_{BA}(b,Q)=\sum_{a,b,j}e^{-S(b,Q)}
[D_{B/b}\circ {\cal C}^{out}_{b/j}] [{\cal
C}_{j/a}^{in}\circ f_{a/A}].
\eeq
In this formula, the ${\cal C}$-functions, 
which are convoluted with the PDFs $f_{a/A}$, or
FFs $D_{B/b}$, arise because 
of collinear radiation along the direction of the
initial or final LO quark. The contributions from the sea of soft
partons that are not associated with either 
of the two jet-like structures
are absorbed into the Sudakov factor $S(b,Q)$. Another part of the
Sudakov factor is used to parameterize the contribution from
non-perturbative physics,
which cannot presently be calculated from first principles.   
The integral over the parameter $\vec b$ is combined 
with the known fixed-order 
perturbative cross-section $\sigma^{pert}_{BA}$. To avoid double
counting, one must subtract the singular part of the perturbative
cross-section $\sigma^{sing}_{BA}$,
which was already included in the $b$-space integral. 

In Fig.\,2 we compare experimental results of the H1 Collaboration 
\cite{desy95108} to
the predictions of the resummation formalism for the $z$-flow
(rescaled transverse energy flow in the $\vg p$~c.m.~frame). The
precise definition of the $z$-flow is
\beq
\frac{d \Sigma_z}{d x \ dQ^2 \ dq_T^2} =
\sum_B \int_{z_{min}}^1 \ z \
\frac{ d\sigma(e A \ra e B X) }{dx \ dz \ dQ^2 \ dq_T^2}\ d z,
\eeq
where $z = (p_a \cdot p_b)/(p_a \cdot q)$ is a variable describing the fragmentation 
of the final parton into
the observed hadrons, and the cross-sections in the limit $q_T \ra 0$
are calculated according to (\ref{resum}).
Using (\ref{qt}), the experimental values 
for $\frac{d \Sigma_z}{d x dQ^2 dq_T^2}$
can be derived from the published pseudorapidity distributions of the
transverse energy flow $\langle E_T \rangle$ in the $\vg p$ c.m.~frame.

\noindent
\parbox{6cm}{
\
\hspace{1\parskip}
As can be seen in Fig.\,2, the results of the resummation 
calculation \cite{NSY} are in acceptable agreement with the data. Note that
the study of the small $q_T$ resummation formalism at HERA has several
advantages, including the possibility to test the resummation formula
at very small values of Bjorken $x$. It will be interesting to
compare the predictions of the resummation formalism for various types
of hadronic final states, since this will allow a test of the
universality of the non-perturbative Sudakov factor. To summarize,
the study of semi-inclusive DIS in the current region of the $\vg
p$~c.m.~frame may provide important insights on the nature of the
soft and collinear parton radiation in the hadronic interactions at
high energies.}
\parbox{6cm}
{
\epsfxsize 6cm
\epsffile{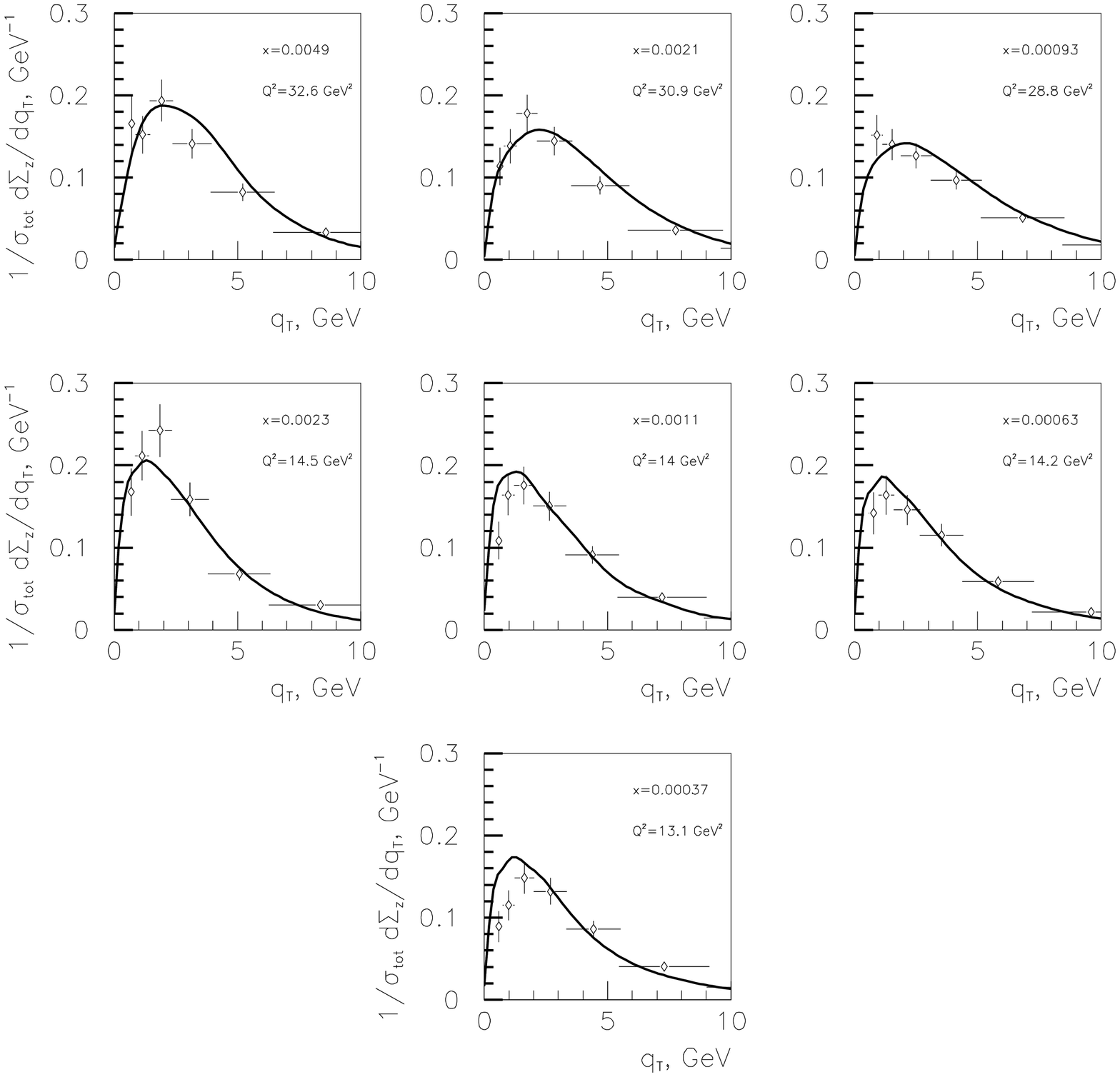}
{\footnotesize Figure 2: 
Comparison of the resummed $z$-flow with the experimental data from 
the H1 Collaboration \cite{desy95108}}
}

\section*{Acknowledgments}
P.\ N.\ thanks the organizers of the DIS2000 Workshop
 for their warm hospitality.
This work was supported in part by the NSF under grant
PHY-9802564.

\end{document}